\documentclass[pre,preprint,amsmath,amssymb,floats,showpacs]{revtex4}

\usepackage{graphicx}

\def \beq{\begin{equation}}
\def \eeq{\end{equation}}
\def \bea{\begin{eqnarray}}
\def \eea{\end{eqnarray}}
\def \karman{$\mathrm{K\acute{a}rm\acute{a}n}$}

\begin{document}

\title{Numerical Investigation of Isolated Crescent Singularity}

\author{Tao Liang}
\affiliation{The James Franck Institute and the Department of Physics, The University of Chicago, 929 East 57th Street, Chicago, IL 60637}

\date{\today}


\vskip 2cm

\begin{abstract}
In this paper we examine numerically the properties, especially the scaling properties, of an isolated crescent singularity similar to that of a developable cone. The desired isolated crescent region is produced by applying six potential forces to an elastic sheet in a controlled way, for which no central pushing force is required. Two types of length scales of the crescent are identified and shown to scale differently with the thickness and the separation of potentials. It is found that in one direction, the width of the crescent scales with both thickness and separation to the 1/2 power. In the other direction, the radius of curvature of the crescent scales with thickness to the 1/3 power and separation to the 2/3 power, in agreement with previous observation for the crescent size of a developable cone. We expect our findings of the double features of the crescent singularity to have importance in understanding the puzzling scaling behavior of the crescent.   
\end{abstract}

\maketitle

\section{Introduction}

In the last several years, there has been a marked interest in the nature of crumpling. Whereas crumpling is a common phenomenon that we encounter every day, it exhibits some of the more intriguing behaviors of modern soft matter physics, such as phase transitions \cite{nelson1}, scaling \cite{alex3}, and energy condensation \cite{kramer1}. Particularly, strong deformations of thin membranes and plates occur in many systems at various length scales, from membranes of polymerized phospholipids \cite{sackmann} to monomolecular layers such as graphite oxide \cite{spector} or molybdenum disulfide \cite{chianelli}, to mountian ridges as a result of strong deformation of earth's tectonic plates \cite{bull}. In intermediate scales, the mechanical properties of thin elastic plates and shells undergoing large deformations are of great importance in engineering of safety structures \cite{stut} and packaging material development \cite{newton}. 

When a thin elastic sheet is confined to a region much smaller than its size, the energetically preferred configurations of the crumpled sheet consist of mostly flat regions bounded by straight ridges or folds that meet at sharp points or vertices. These ridges and vertices constitute the two types of singular structures that appear on a crumpled sheet. Energies are mostly condensed into a network of such singularities. These singularities, as a result of the stress focusing, have recently been the subject of several investigations[10-28]. Related phenomena, such as thin-film blistering[29-31], thin viscous sheet \cite{boud2}, thin-film actuators \cite{bha}, molecular sheets[11,34], and the generalization of crumpling to higher dimensions [3,19-21] have also received attention. In particular, the properties of ridges have been studied extensively by Witten and collaborators. It was discovered that the structure of these ridges could be accounted for using linear elasticity theory, valid in the limit that the ridge length $X$ is much greater than the sheet thickness $h$ [2,11-13]. Witten and Li used a scaling argument to predict that the ridge possesses a characteristic radius of curvature given by $R \sim h^{1/3}X^{2/3}$ and the total elastic energy scales as $E \sim \kappa (X/h)^{1/3}$, where $\kappa$ is the bending modulus of the sheet \cite{witten}. Lobkovsky \textit{et al} verified these scaling laws through an asymptotic analysis of the von \karman~equations for a thin sheet and through numerical simulations [12,13]. DiDonna \textit{et al} investigated the buckling of a single stretching ridge under external forcing and demonstrated the scaling laws in the buckling instability \cite{brian2}. 

On the other hand, considerable work has been done to understand the properties of the vertex singularities. Developable cones ($d$ cone) are found to be a particular solution of the von \karman~equations describing large deformations of thin plates \cite{ben-amar}. Cerda and Mahadevan analyzed the geometry and elasticity of a single developable cone formed by pushing an elastic sheet against a circular ring (see Fig.~\ref{dcone}), in the limit of pure bending \cite{maha-new}. They obtained the shape of the $d$ cone by minimization of bending energy and showed that the aperture angle of the buckled part has a universal value of 139 degrees. Experimentally, several different groups have studied the mechanical properties of $d$ cone [17,22-24,42]. The dynamic interactions of two $d$ cones in a simplified geometry has especially attracted experimental interest \cite{boud}. Most of this work assumed that $d$ cone has a sharp vertex point at its tip. However, that is only true in the unstretchable limit. For a physical sheet with finite thickness, the vertex point expands to a core region of finite size in which energetically expensive stretching is localized. The core size is governed by the competition of the bending and stretching energies. As a result of stress focusing in the core region, crescent-like shapes appear where bending stresses are big.  

The problem of interest here is concerned with the size of the core region. We want to know whether there is scaling behavior of the core size and determine any scaling exponents. Cerda \textit{et al} proposed that, for a single $d$ cone, the core size $R_c$ scales as $R_c \sim h^{1/3} R^{2/3}$, where $h$ is the thickness and $R$ is the size of the confining ring (see Fig.~\ref{dcone}) \cite{cerda-nature}. This scaling proposition was verified by numerical simulations using two different methods, where elastic sheet is modelled by a triangular lattice of springs \cite{last}. This result is surprising because stretching energy is supposed to be concentrated in the core region, so the core should not be able to know about the length of the outside confining ring. In this work, we tackle this puzzle by studying an isolated core region in the sheet. To achieve this geometry, instead of pushing at the tip of the $d$ cone against a confining ring, we apply constraints on the side of the sheet to form a crescent-like region in the middle of the sheet. The details about the implementation are to be explained in Section III. The resulted geometry is similar to a Cerda-Mahadevan $d$ cone in that it exhibits crescent-like region. However, it is fundamentally different from Cerda-Mahadevan $d$ cone: it doesn't have a pushing tip and confining ring, the crescent region becomes much greater and we don't have a large outer region where bending is dominant. These features of the shape make it possible for us to examine the properties, especially the scaling behavior, of the crescent region in greater detail.

The paper is organized as follows. In Section II, we describe the energies and forces that give rise to the crescent singularity, and state some of the observed $d$ cone properties. In Section III, we demonstrate our numerical models and show the way of producing the desired shape in details. In Section IV, we give the detailed description of the observed scaling laws and compare them with the ring ridge (explained later) and straight ridge scalings, as well as $d$ cone scaling. Finally, the limitations and implications from our findings are discussed in Section V.

\section{Developable Cone}
We begin by stating the connection between the deformation of the sheet and its elastic energy. In equilibrium, the sheet assumes a conformation that minimizes the elastic energy. Two forms of energies must be considered, bending energy $B$ and stretching energy $S$. The bending energy density is proportional to the square of the total curvature $C(r)$ \cite{mansfield}. This $C(r)$ defined as the trace of the curvature tensor, is sometimes called the mean curvature \cite{nelson}. The constant of proportionality is called the bending modulus $\kappa$. Thus
\beq
B=\frac{1}{2} \kappa \int dA~C(r)^2~~, 
\eeq
where $\int dA$ denotes the integral over the surface. In general, there is a second form of bending energy, proportional to the average Gaussian curvature. It is shown that this Gaussian curvature energy is negligible for the $d$ cone system \cite{last} despite the potential contribution from the boundary. The stretching energy density is proportional to the square of the strain tensor $\gamma_{\alpha\beta}$. For an isotropic material, the stretching energy can be expressed
\beq
S=\frac{Yh}{2(1 - \nu^2)}\int dA \left[\nu~(Tr~{\bf\gamma})^2 + (1-\nu)~Tr({\bf\gamma}^2)\right]
\eeq
where $Y$ is Young's modulus and $\nu$ is Poisson's ratio \cite{landau}. They are related to bending modulus through $\kappa = Yh^3/(12(1-\nu^2))$. 

The relation that strain tensor $\gamma_{\alpha\beta}$ and curvature tensor $C_{\alpha\beta}$ must satisfy in order to define a surface is the Gauss {\it Theorema Egregium} \cite{milman}. It expresses the Gaussian curvature $K=det(C_{\alpha\beta})$, the determinant of the curvature tensor, in terms of the strain tensor
\beq
K = \partial_\alpha \partial_\beta \gamma_{\alpha\beta} - \bigtriangledown^2 Tr(\gamma_{\alpha\beta})~~,
\eeq
where summation over the repeated indices is implied. This equation geometrically captures the intuitive notion that nonzero Gaussian curvature (the sheet curves in both directions) must cause the sheet to strain. Historically it is called ``geometrical von \karman~equation''. 

On the other hand, the actual strain and curvature fields are those which minimize the $B+S$. The variational minimization amounts to a statement that the normal forces on each element must balance. This statement is known as the ``force von \karman~equation''
\beq
\partial_\alpha \partial_\beta M_{\alpha\beta} = \sigma_{\alpha\beta}C_{\alpha\beta} + P~~,
\eeq
where $M_{\alpha\beta}$ are the torques per unit length, $\sigma_{\alpha\beta}$ are in-plane stresses, and $P$ is external force per unit area of the sheet.

\begin{figure}[!htb]
\begin{center}
\includegraphics[width=\textwidth]{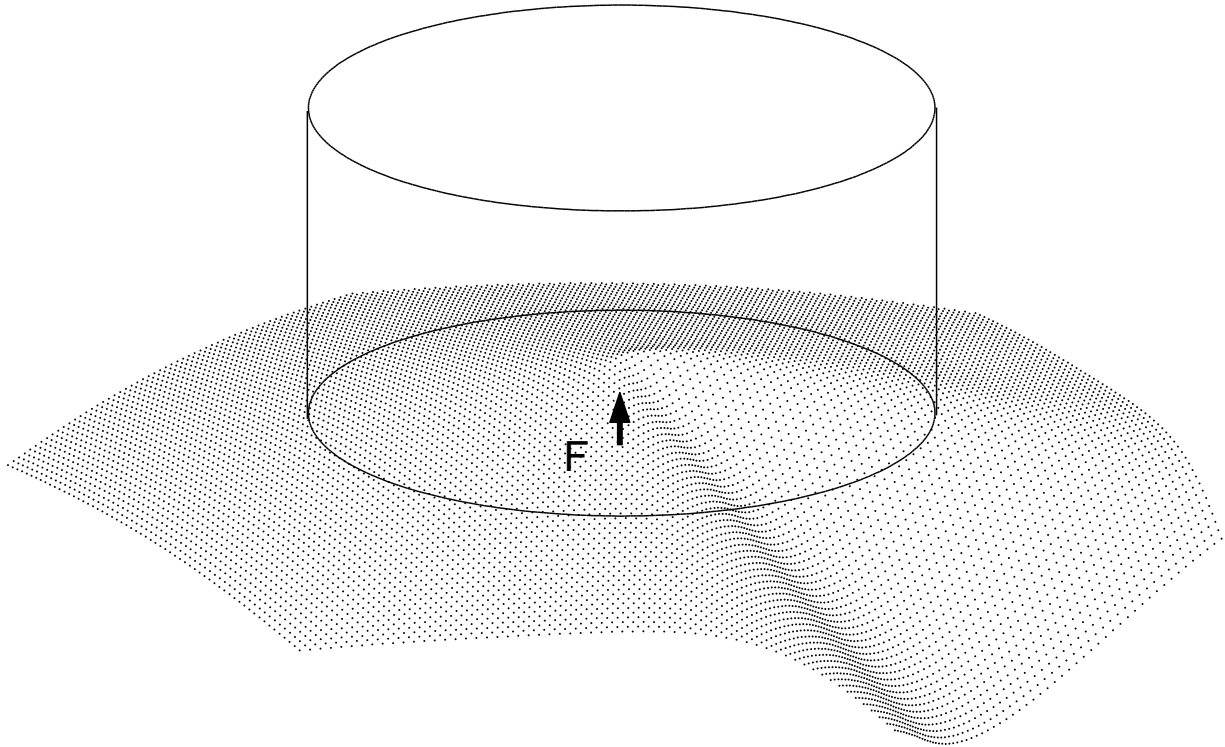}
\caption{A developable cone appears when we push the center of an elastic sheet against a circular ring from below with force $F$. The radius of the confining ring is denoted by $R$.}
\label{dcone}
\end{center}
\end{figure}

After the above general descriptions about the large deformation of an elastic sheet, we now consider a single conical vertex studied by Cerda \textit{et al} \cite{cerda-nature}. One realization is to push the center of a circular elastic sheet of radius $R_p$ against a circular ring of radius $R$, as illustrated in Fig.~\ref{dcone}. The pushing force is perpendicular to the plane of the ring. This is the simplest volume-restricting deformation of the sheet. Due to the constraint of unstretchability, the sheet deforms into a non-axisymmetric conical surface which is only in partial contact with the ring. In the limit that thickness $h$ of the sheet goes to zero, since bending modulus ($\sim h^3$) vanishes faster than the stretching modulus ($\sim h$), there would be pure bending over the sheet and Gaussian curvature would be zero everywhere except at the tip. Mathematically, such a conical surface is called perfectly developable cone \cite{struik}. It is deformed from, or applied to, a plane without changing distances. In this limit, some models about the shape of the $d$ cone have been proposed [16-18,27]. These models only give the outer region solutions of $d$ cone shape, in the sense that they do not consider the stretching energy that is inevitable on a real sheet with finite thickness. A real sheet must stretch near the tip, because otherwise, the curvature at the tip would be divergent, since curvature goes as $1/r$, where $r$ is the distance to the tip, thus causing divergent energy. Therefore, it is the finite thickness that causes the sheet to stretch greatly in a small region near the tip. This small region is called the core region. It is where energetically expensive stretching is localized and its size is governed by the competition of the bending and stretching energies of the whole cone. In addition, as shown in Ref.[25], finite thickness also causes a small amount of strains in the outer region. It is noted that while this geometry is a special realization of the general concept of $d$ cone as defined by Ben Amar and Pomeau \cite{ben-amar}, it is representative of the vextex singularities on a crumpled sheet and we herein refer this geometry as $d$ cone.  

A few interesting properties have been revealed for this simple geometry. It has been found both theoretically and numerically that the normal forces exerted by the ring on the sheet have a singular term at the take-off points, where the sheet begins to leave the ring [25,27]. The ratio of the normal force from the singular term to that from the constant term is found to be about 0.70. Besides, it is observed numerically that the constant normal forces produce a striking effect: they induce a radial curvature on the sheet near the supporting ring and cause the mean curvature to vanish there \cite{last1}. However, the observed puzzling scaling law about the core size has yet to be understood. In order to study the core region in greater details, we need to isolate the core region from the outer region. It is desirable to form the core region without exerting the central pushing force. For this purpose, a six-point pattern of forces is applied to the sheet; this is discussed in the next section.

\section{Numerical Methods}
An elastic sheet is modelled by a triangular lattice of springs of un-stretched length $a$ and spring constant $k$, after Seung and Nelson \cite{nelson}. Bending rigidity is introduced by assigning an energy of $J(1-\hat{n}_1 \cdot \hat{n}_2)$ to every pair of adjacent triangles with normals $\hat{n}_1$ and $\hat{n}_2$. When strains are small compared to unity and radii of curvature are large compared to the lattice spacing $a$, this model bends and stretches like an elastic sheet of thickness $h=a\sqrt{8J/k}$ made of isotropic, homogeneous material with bending modulus $\kappa = J\sqrt{3}/2$, Young's modulus $Y=2ka/h\sqrt{3}$ and Poisson's ratio $\nu=1/3$. Lattice spacing $a$ is set to be 1. The shape of the sheet in our simulation is a regular hexagon of side length $R_p$. The typical value of $R_p$ is $60a$.

\begin{figure}[!htb]
\begin{center}
\includegraphics[width=\textwidth]{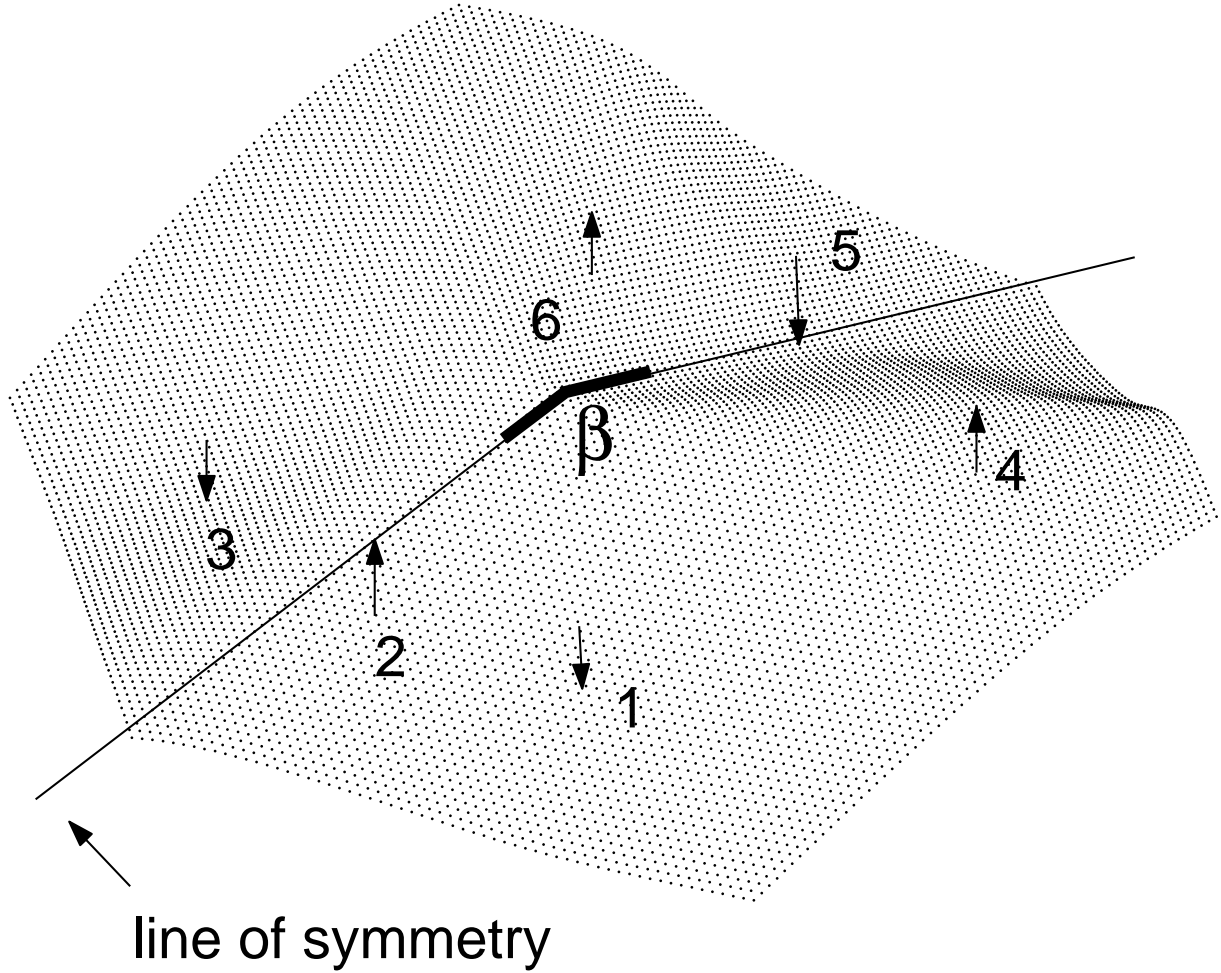}
\caption{A typical shape of the simulated crescent region, formed under the exertion of six forces in a controlled way. The arrows indicate the directions of the applied forces, as well as their locations. The numbering of the forces is the same as that shown in Fig.~\ref{sketch}. For this shape, the sheet has thickness $h=0.102a$, and two sets of potentials are separated by $60a$, where $a$ is the lattice spacing in the numerical model. The deformation angle $\beta$ shown in the figure is to be defined in Section IV.A. For this shape, $\beta = 2.763$ rad.}
\label{core}
\end{center}
\end{figure}

To create a crescent region in the sheet, we need to apply forces to the sheet. Unlike $d$ cone, we do not apply a pushing force to the center of the sheet against a confining ring. Instead, we apply six forces to the sheet in a special manner to produce the desired crescent region. Fig.~\ref{core} shows a typical shape of the produced crescent region. The arrows in the figure show the directions of the applied forces. The details of implementation are as follows.

\begin{figure}[!htb]
\begin{center}
\includegraphics[width=0.9\textwidth]{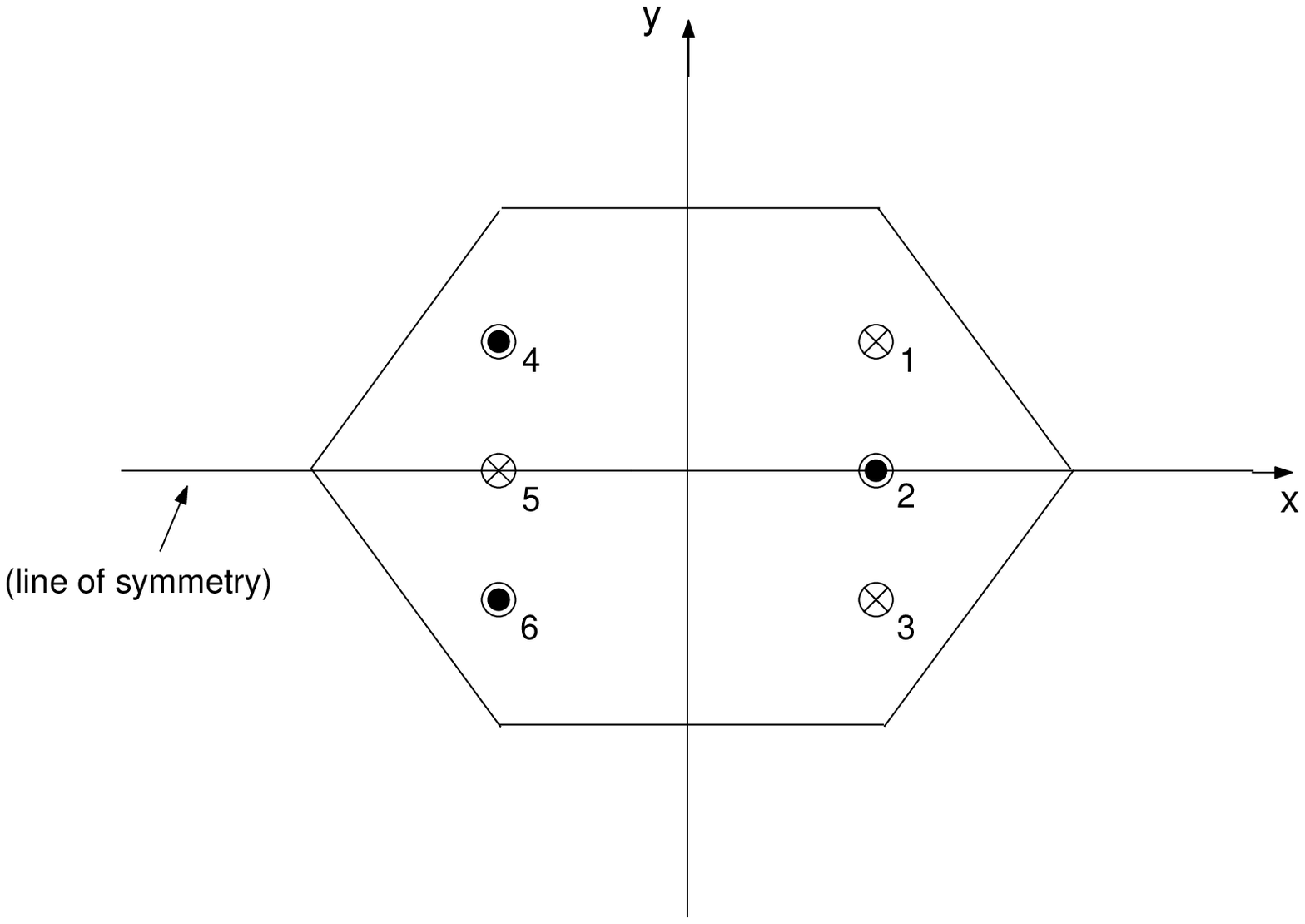}
\caption{The sheet is in its flat state and we apply six potentials to the sheet. The center of the sheet is taken to be the origin. $z = x \times y$. There are two sets of potentials: \{1,2,3\} and \{4,5,6\}. They are symmetric with respect to the $y$ axis. Big circles denote the positions of the potential centers. The dots within the circles mean the potentials push the sheet in the positive $z$ direction (out of the page); the crosses within the circles mean the potentials push sheet in the negative $z$ direction (into the page).}
\label{sketch}
\end{center}
\end{figure}

First, each force is implemented by a spherically symmetric potential of the form $U=\sum A/[((x_i-x_0)^2+(y_i-y_0)^2+(z_i-z_0)^2)^4+\xi^8]$, where $A$, $\xi$ are constants and the summation is over all lattice points with coordinates $(x_i,y_i,z_i)$. $(x_0,y_0,z_0)$ are the coordinates of the center of the potential. We choose the range $\xi$ of the potential to be one lattice spacing. This potential is repulsive and the pushing force it produces decays rapidly once the lattice points go away from the center of the potential. In the simulation, we apply two sets of potentials to the sheet, with each set consisting of three potentials, as shown in Fig.~\ref{sketch}. In this figure, the big circles denote the positions of the potential centers. All the six potential centers lie in the $x-y$ plane. The first set \{1,2,3\} is symmetric to the second set \{6,5,4\} with respect to the center of the sheet (origin). Specifically, the potential centers 1 and 6 are symmetric to each other with respect to the origin. So are the centers 2 and 5, and 3 and 4. The potential centers within each set lie on the same straight line. For each set, the amplitude of the potential in the middle is twice the amplitude of the potentials on two sides, and the sheet adjusts itself to maintain equilibrium. We define the `separation' to be half of the distance between the two straight lines of the two sets. Typical values of coordinates of the six potential centers are $(R_p/2,R_p \sqrt{3}/4,0)$, $(R_p/2,0,0)$, $(R_p/2,-R_p\sqrt{3}/4,0)$, $(-R_p/2,R_p\sqrt{3}/4,0)$, $(-R_p/2,0,0)$, $(-R_p/2,-R_p\sqrt{3}/4,0)$, where $R_p$ is the side length.

Second, to control the direction of the forces exerted by the six potential forces, we set the initial states of the sheet such that forces 2, 4 and 6 are acting upwardly and forces 1,3 and 5 are acting downwardly, as shown in Fig.~\ref{core} and Fig.~\ref{sketch}. We realize this by moving the lattice points near the potential centers up or down by certain distance from the $x-y$ plane. Since the potential is repulsive, this assures that force will act in the direction we want it to be. For example, we move the lattice points near potential center 2 upward from the $x-y$ plane, so that potential 2 will push the sheet in the positive $z$ direction. The six force constraints exerted in this way will be likely to produce the desired crescent region near the center of the sheet.  

The conjugate gradient algorithm \cite{recipe} is used to minimize the total elastic and potential energy of the system as a function of the coordinates of all lattice points. This lattice model behaves like a continuum material provided that the curvatures are everywhere much smaller than $1/a$. 
 
However, after running the simulations as described above, we found that the sheet tends to slide out of the contraints of potentials. This is understandable: the conjugate gradient algorithm looks for the configurations of the sheet with the minimum energy, and thus the sheet always tries to escape the potential whenever it is permitted by the algorithm. To remedy this problem, we impose symmetry on the sheet. Specifically, as shown in Fig.~\ref{sketch}, we use $x-z$ plane as the plane of symmetry: the lattice points with positive $y$ value are the `master' points and the lattice points with negative $y$ values are the corresponding `slave' points. Each slave point is just an image of the master point with respect to the plane of symmetry. The points lying on the plane (with $y=0$) are also master points but they have no slave points. Now the conjugate gradient algorithm is used to minimize the total elastic and potential energy with respect to the coordinates of master points only. In the minimization procedure, the coordinates of the slave points are determined from the master points and the total energy is calculated for the whole sheet. After imposing this symmetry, the sheet no longer slides out of the potential constraints, because otherwise, doing so would increase the energy during the minimization process, which is prohibited by the algorithm.  

As a last note for this section, it is noticed that this lattice model of elastic sheet has been used to study both the ridge and the point-like singularities in crumpled sheets [12-15,25]. The accuracy of it has been tested in various ways. Using this model, Lobkovsky \cite{alex2} numerically verified the ``virial theorem'' for ridges that bending energy is five times the stretching energy, for asymptotically thin sheets. In the study of $d$-cone, the ratio of the normal force from the $\delta$-function term to that from the angle-independent term is found to be 0.70 numerically \cite{last}, compared well with the theoretical prediction 0.69 \cite{maha-new}. Also, the azimuthal profiles of curvature are in reasonably good agreement with the theoretical prediction as deformation of $d$-cone goes to zero \cite{last}. Moreover, we tested our program by setting the lattice at different initial states and letting the program look for minimized energy state. We found the lattice converges to the same state, with agreements of pushing force and energies better than one percent for fixed deflection. In addition, we test our model by changing the lattice spacing and showing our results are independent of the lattice spacing, as we will see in the next section.

\section{Findings}

\subsection{Two Characteristic Length Scales}
The produced crescent region has two characteristic length scales: the radius of curvature of the crescent that characterizes the crescent/core size, and the width of crescent that represents the size of the band over which stretching is concentrated. The former is the core size as referred in Section I, and the latter is the counterpart to the width of a straight ridge. We want to study how these two length scales change with other length scales of the system. 

To obtain the width of the crescent region, we first take a look at the curvature profile along the line of symmetry. This line of symmetry is perpendicular to the crescent at its tip. We determine the curvatures approximately from each triangle in the sheet. For this measurement, we take the curvature tensor to be constant across each triangle. We calculate it using the relative heights of the six vertices of the three triangles that share sides with the given triangle \cite{brian}. The six relative heights $w_i$ normal to the triangle surface are fit to a function of the form
\beq
w_i=b_1+b_2 u_i+b_3 v_i +b_4u_i^2 +b_5 u_i v_i+b_6 v_i^2, ~~i=1, \ldots, 6
\eeq
where \{$u_i,v_i,w_i$\} are coordinates of the vertices in a local coordinate system that has $w$ axis perpendicular to the surface of the given triangle. This choice of local coordinate system ensures that $b_2$ and $b_3$ are negligible so that curvature tensors can be determined only from the coefficients of quadratic terms. In practice, our numerical findings do show that the values of $b_2$ and $b_3$ are on the order of $10^{-2}$ or lower \cite{last}. Therefore, curvature tensors follow immediately from the identification $C_{uu}=2 \times b_4, C_{vv}=2 \times b_6, C_{uv}=b_5$. Futhermore, in our calculation, we choose the directions of local coordinate system in such a way that either $u$ or $v$ axis lies along the line of symmetry. Under such a condition, the cross term $C_{uv}$ is close to zero, and $C_{uu}$, $C_{vv}$ represent the two principal curvature components.   

\begin{figure}[!htb]
\begin{center}
\includegraphics[width=\textwidth]{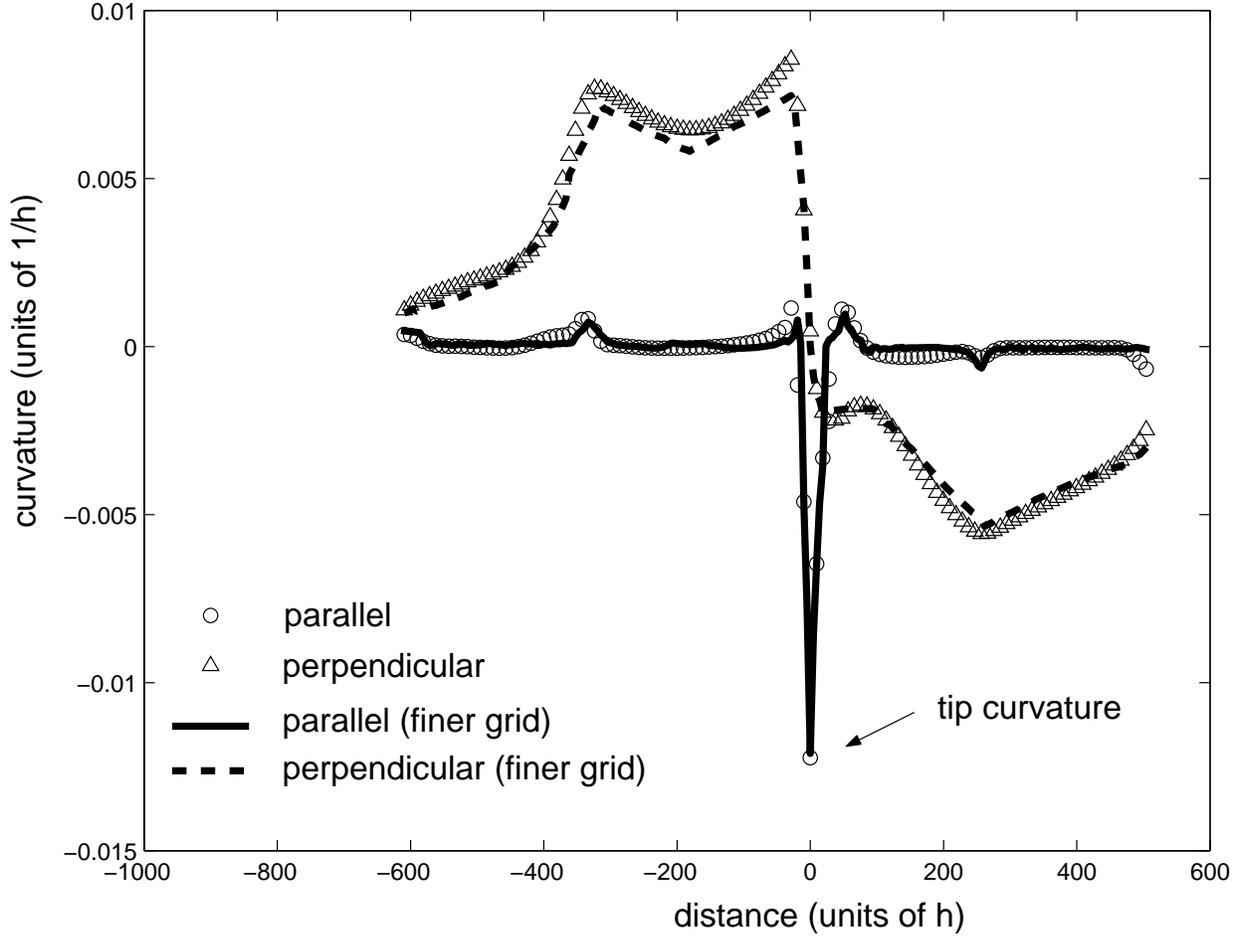}
\caption{Curvature profiles along the line of symmetry. Open symbols are for the shape shown in Fig.~\ref{core}. Solid and dashed lines are for shape simulated with half lattice spacing $a/2$ and finer lattice grid. The distance is measured from the tip of crescent and in terms of $h$. Curvature is measured in terms of $1/h$. The distance between two adjacent data points is approximately one lattice spacing. The circles and solid line denote the principal curvature parallel to the line of symmetry, and the triangles and dashed line denote the principal curvature perpendicular to the line of symmetry. The maximal negative curvature represents the curvature at the tip. The sheet has thickness $h=0.102a$, side length $60a$ and deformation angle $\beta= 2.76$ rad for both shapes. The two sets of potentials are separated by $60a$.}
\label{curv_prof}
\end{center}
\end{figure}

Fig.~\ref{curv_prof} gives typical plots of the curvature profiles along the line of symmetry. The distance is measured from the tip of the crescent and is taken to be negative in the concave region (such as where force 5 points in Fig.~\ref{core}). It's measured in terms of $h$. The curvature is measured in units of $1/h$ and taken be to negative in the convex region (such as where force 2 points to in Fig.~\ref{core}). The distance between two adjacent data points is approximately one lattice spacing. We first focus on the open symbols in the figure. Data denoted by open circles is the principal curvature parallel to the line of symmetry, and the data denoted by open triangles is the principal curvature in the direction perpendicular to the line of symmetry. The shape of the curves is in qualitative agreement with our analysis below. The principal curvature parallel to the line of symmetry is mostly zero in the region away from the crescent, and takes negative values within the crescent region. It reaches the maximum negative value at the position of the tip, where the sheet is most curved in the direction of line of symmetry, as pointed out in the figure. This maximum value characterizes the width of the crescent region. Curvature parallel to the line of symmetry also has a small peak and a small dip at about half way into the negative and positive distances, respectively. These are caused by the potential forces, as there are are small deformations in the sheet near the potential centers. In addition, the curvature perpendicular to the line of symmetry is effectively zero at the tip, and as expected, it changes signs as we go from concave region to the convex region. 

As a further test of our numerical model, we reduce the lattice spacing by half but keep all other length scales including thickness fixed. Therefore, we have a finer lattice grid for the same sheet, and we can compare the curvature profile obtained from this finer grid with the one from the original grid. Practically, this transformation is equivalent to fixing lattice spacing, doubling the thickness, the separation of potentials and the size of the sheet. The solid and dashed lines in Fig.~\ref{curv_prof} show the curvature profiles for the new shape with finer grid. It is readily seen that the parallel principal curvature profile (solid line) follows similar the shape as that of our original grid (circles), and the tip curvature is almost the same, indicating that it is independent of lattice spacing. The perpendicular principal curvature profile (dashed line) looks also similar to the original grid profile (triangles), but it does not have the sharp kinks exhibited in the original grid profile. This is attributed to the lattice effect, since we have more lattice points in the finer grid. 

\begin{figure}[!htb]
\begin{center}
\includegraphics[width=0.8\textwidth]{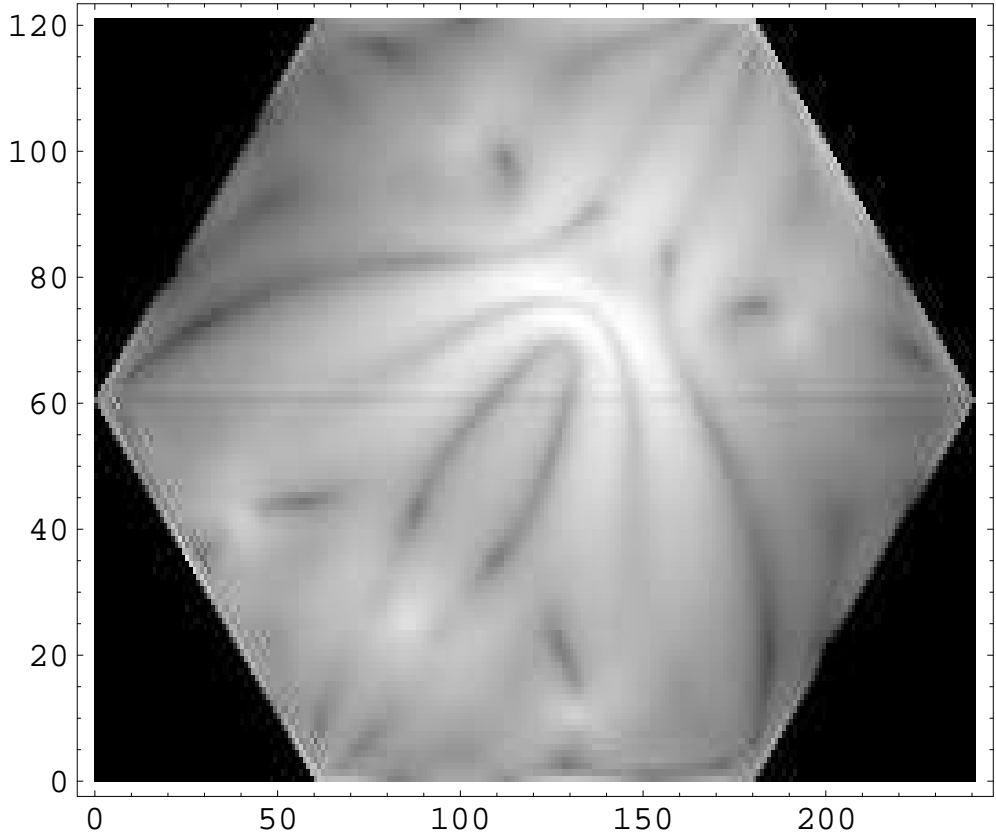}
\caption{Density plot of logarithmic stretching energy over the surface of the sheet, for the shape described in Fig.~\ref{core}. The line of symmetry runs from the lower left corner to the upper right corner. On the gray scale, brighter region has higher energy. The top 3\% lattice points with the highest stretching energies are concentrated in the crescent region and they represent nearly 75\% of the total stretching energy.}
\label{denplot}
\end{center}
\end{figure}

Having obtained the width, we now turn to the other length scale --- the crescent size. We want to measure the radius of curvature at the tip of the crescent region. For this purpose, we use the stretching profile over the surface of the sheet. We obtain the stretching energy associated with each of the lattice point from the sum of the stretching energies of the springs connected to it. Fig.~\ref{denplot} displays a typical density plot of the stretching energy over the sheet as projected onto the $x-y$ plane. We take the logarithm of the energy to reduce the contrast so that we have a more clear view of the energy distribution near the tip. It is readily seen that the line of symmetry runs from the lower left corner to the upper right corner. As expected, the stretching energy is localized in the crescent region. We find that the top 3\% lattice points with the highest stretching energies are concentrated in the crescent region and they represent nearly 75\% of the total stretching energy.  

\begin{figure}[!htb]
\begin{center}
\includegraphics[width=0.8\textwidth, bb=0pt 0pt 380pt 275pt]{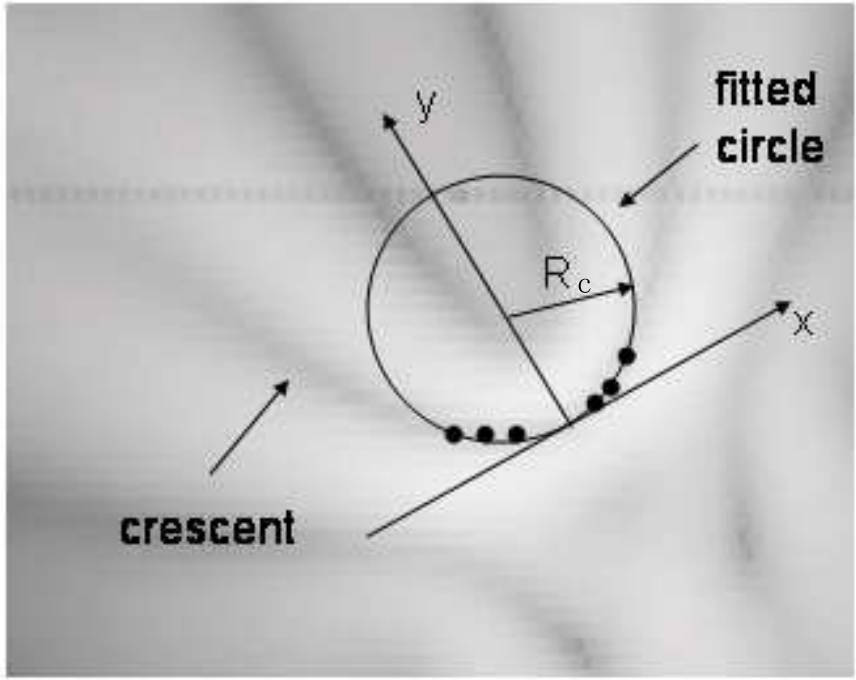}
\caption{Fitting of a circle to six points extracted from the crescent line in the density plot of stretching energy as displayed in Fig.~\ref{denplot}. The radius of the fitted circle is the crescent size $R_c$.}
\label{fit}
\end{center}
\end{figure}

The radius of curvature of the crescent can be measured from the dark curve lying against the bright region near the center. As shown in Fig.~\ref{fit}, we take six points on the curve near the tip, three on each side, and we fit a circle to these six points. Suppose the radius of curvature is $R_c$. In the local coordinate system, the coordinates of six points are ($x_i,y_i$) for $i=1,\cdots,6$. We have $x^2+(y-R_c)^2=R_c^2$, so $x^2+y^2 = (2R_c) y$. We use least square techniques to find the best estimate of $R_c$. It is $\hat{R_c} = \sum (x_i^2+y_i^2)y_i/(2\sum_i y_i^2)$, where summation is over 6 points.  

We now define the deformation angle. As shown in Fig.~\ref{core}, deformation of the crescent region is characterized by the angle made at the tip by the two segments of the line of symmetry in convex and concave parts of the sheet. We call it angle $\beta$. The larger the deformation, the smaller the $\beta$. It is noted that the calculation of $\beta$ is straightfoward. 

\subsection{Thickness Scaling}

\begin{figure}[!htb]
\begin{center}
\includegraphics[width=0.9\textwidth]{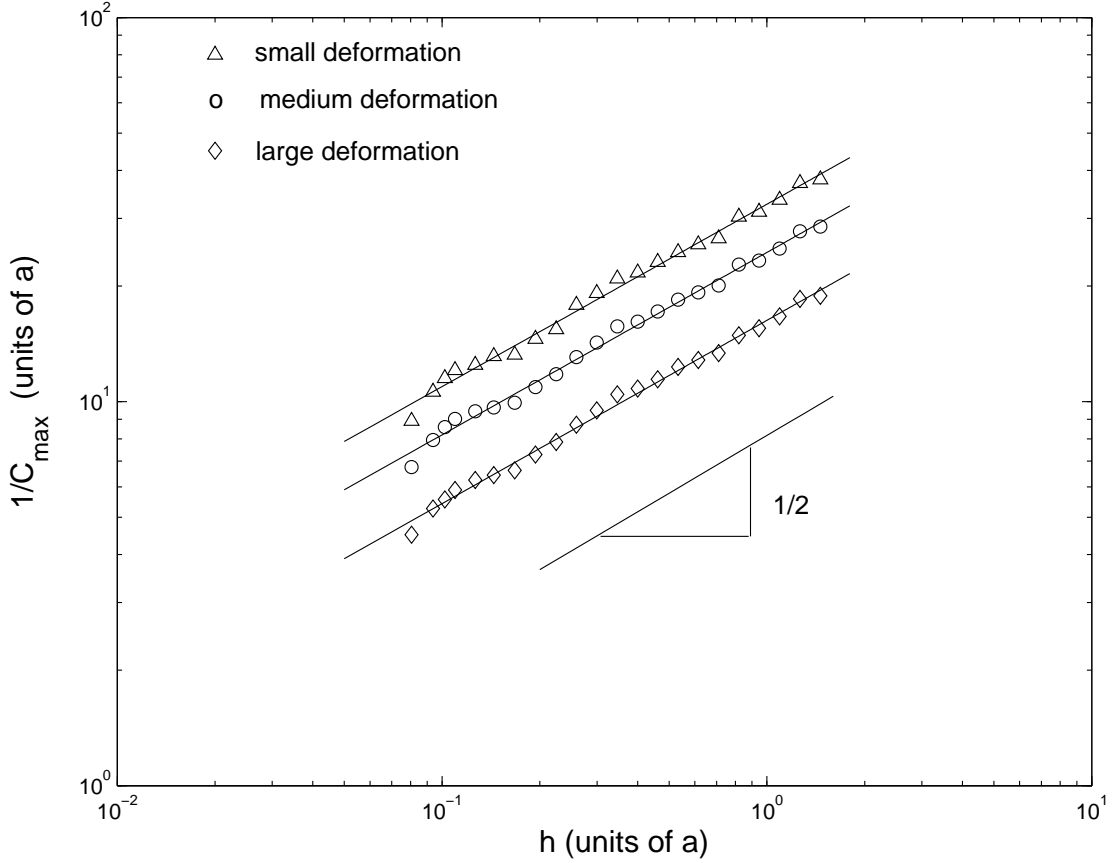}
\caption{The plots of width versus thickness for three different deformations. The plots are on log-log scale. Width is calculated from $1/C_{\mathrm{max}}$, where $C_{\mathrm{max}}$ is the curvature at the tip, as determined from the curvature profile. Triangles denote data for small deformation with $\beta=2.926$; circles denote data for medium deformation with $\beta=2.743$; diamonds denote data for large deformation with $\beta=2.635$. The slopes of three fitted lines are 0.474, 0.475 and 0.477, respectively.}
\label{width_thickness}
\end{center}
\end{figure}

As mentioned above, using the maximum negative value of the principal curvature parallel to the line of symmetry, we can obtain the width of the crescent. We want to study how width of crescent changes with the thickness. We change thickness and measure the corresponding width from the curvature profile along the line of symmetry. We change thickness by changing the bending coefficient $J$ and keeping the spring constant $k$ fixed. We fix the separation of the potentials at $30a$, that is, the distances between the two sets of potentials is $60a$. To fix the deformation while changing thickness, we adjust the amplitude of the potential $A \sim h^2$ accordingly, because bending modulus $\kappa$ goes as $h^2$ as $k$ is fixed. In practice, this scheme works well --- we find that the deformation angle $\beta$ changes no more three percents as we change the thickness. Fig.~\ref{width_thickness} shows how width changes with thickness, for three different deformations. The triangles, circles and diamonds denote the data for small, medium and large deformations. The values of deformation angle $\beta$ in three runs are 2.926, 2.743 and 2.635 rad. The slopes of the three fitted lines are 0.474, 0.475 and 0.477, respectively, all close to 0.5. This suggests that width goes as $h^{1/2}$. 

\begin{figure}[!htb]
\begin{center}
\includegraphics[width=0.9\textwidth]{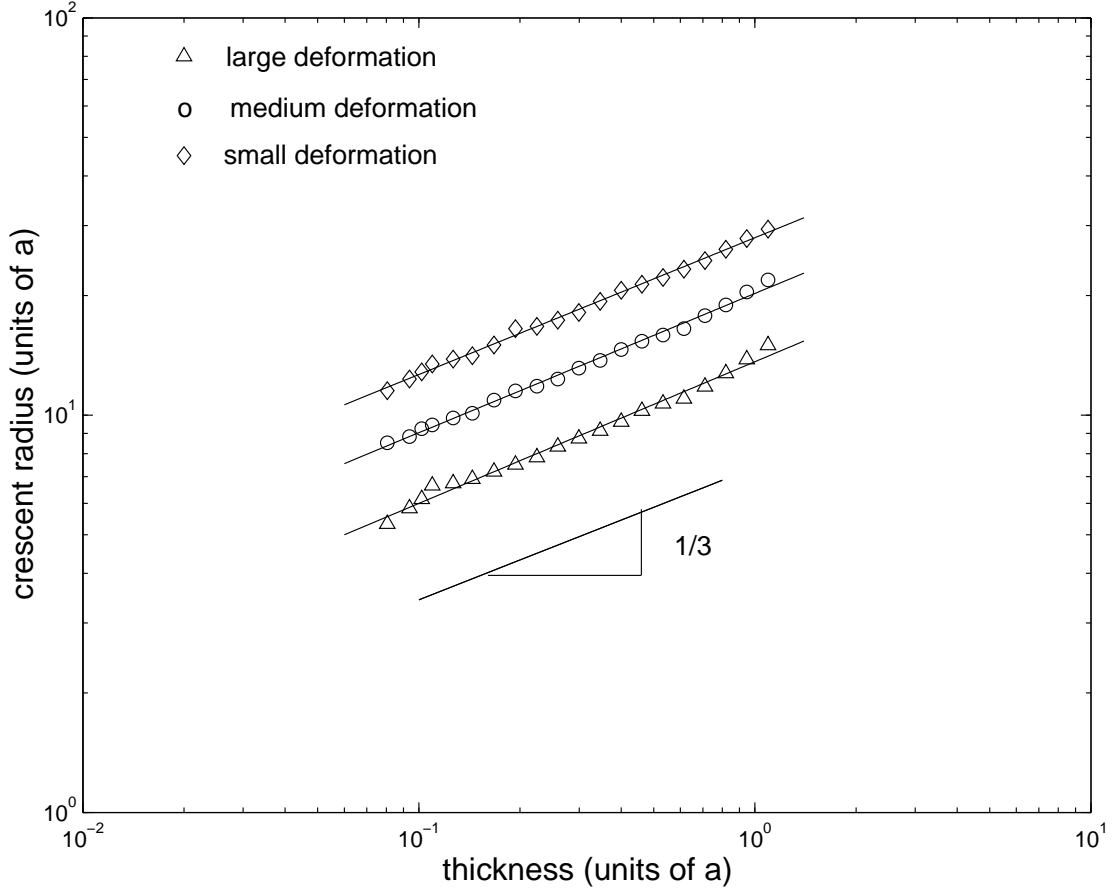}
\caption{The plots of crescent size versus thickness for three different deformations, on log-log scale. Both lengths are in units of lattice spacing. The crescent sizes are measured from the same simulated shapes as for Fig.~\ref{width_thickness}. The fitted values of slope are 0.344, 0.351 and 0.357.}
\label{crescent_thickness}
\end{center}
\end{figure}

As discussed in Section IV.A, the crescent size is calculated by fitting circle to the points extracted from the density plot of stretching energy. Fig.~\ref{crescent_thickness} shows crescent size as a function of the thickness, for three different deformations. We follow the same scheme to keep deformation fixed as we did before. The fitted values of the slope is 0.344, 0.351 and 0.357, for small, medium and large deformations, respectively. This suggests that crescent size goes as $h^{1/3}$, which is the same as what we observed for crescent size in a $d$ cone \cite{last}. It is noted that width of a straight ridge also depends on thickness to the 1/3 power.  

\subsection{Separation of Potentials}
We now study the effect of changing separations between the two sets of potentials. As mentioned in Section III, the separation is denoted by half of the distance between them. As we adjust the positions of one set of potentials, we adjust the positions of the other set accordingly, such that they maintain their symmetry with respect to the $y$ axis. The separation is a measure of the `force arms' by which we constrain the sheet. We change the positions of potentials in the following way: we move all three potentials within one set together, parallel to the $x$ axis. That is, we keep the $y$ coordinates of all the potential centers fixed, and just change their $x$ coordinates. To keep the deformation of the sheet fixed during the change of separation, we adjust the amplitude of the potentials such that it is inversely proportional to the separation. By this way we can make the total torque exerted on the sheet fixed. Once again, we found that the deformation angle changes no more than five percents as we change the separation.

\begin{figure}[!htb]
\begin{center}
\includegraphics[width=0.8\textwidth]{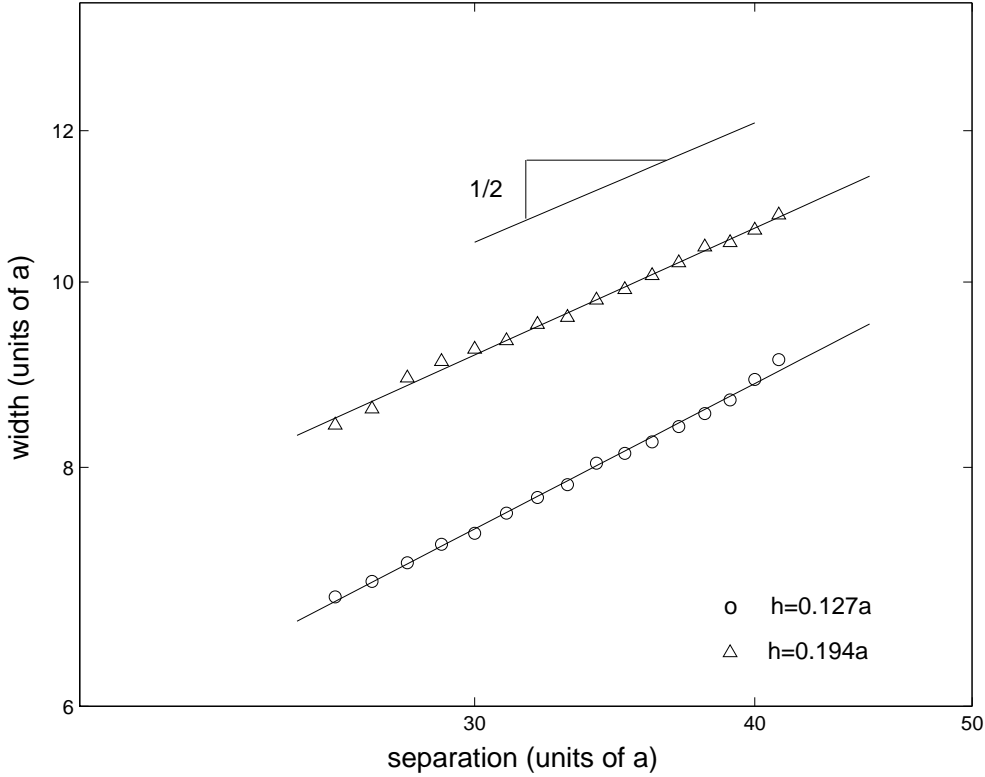}
\caption{Plots of width versus separation, for two different thicknesses. Thickness is fixed at $h=0.127a$ and $h=0.194a$. The fitted lines have slope 0.609 and 0.531, respectively. The deformation angle is kept fixed at $\beta=2.74$ rad in both runs.}
\label{width_sep}
\end{center}
\end{figure}

Fig.~\ref{width_sep} shows how width changes with the separation for two different thicknesses. In practice, we can not change separation over a large range, so it just runs from $26a$ to $41a$ for our sheet of side length $60a$. The reason that we do not have a large range of separation is that the sheet tends to escape the constraints of potentials as the separation becomes too large, and in our sequential minimization process, the forces are likely to change direction as the separation gets too small. Within this range of separation, we observe a scaling relationship between width and separation, as shown in Fig.~\ref{width_sep}. The fitted values of slope are 0.609 and 0.531, for thickness $h=0.127a$ and $h=0.194a$, respectively. It is noted that they are not far from 1/2. 

\begin{figure}[!htb]
\begin{center}
\includegraphics[width=0.8\textwidth]{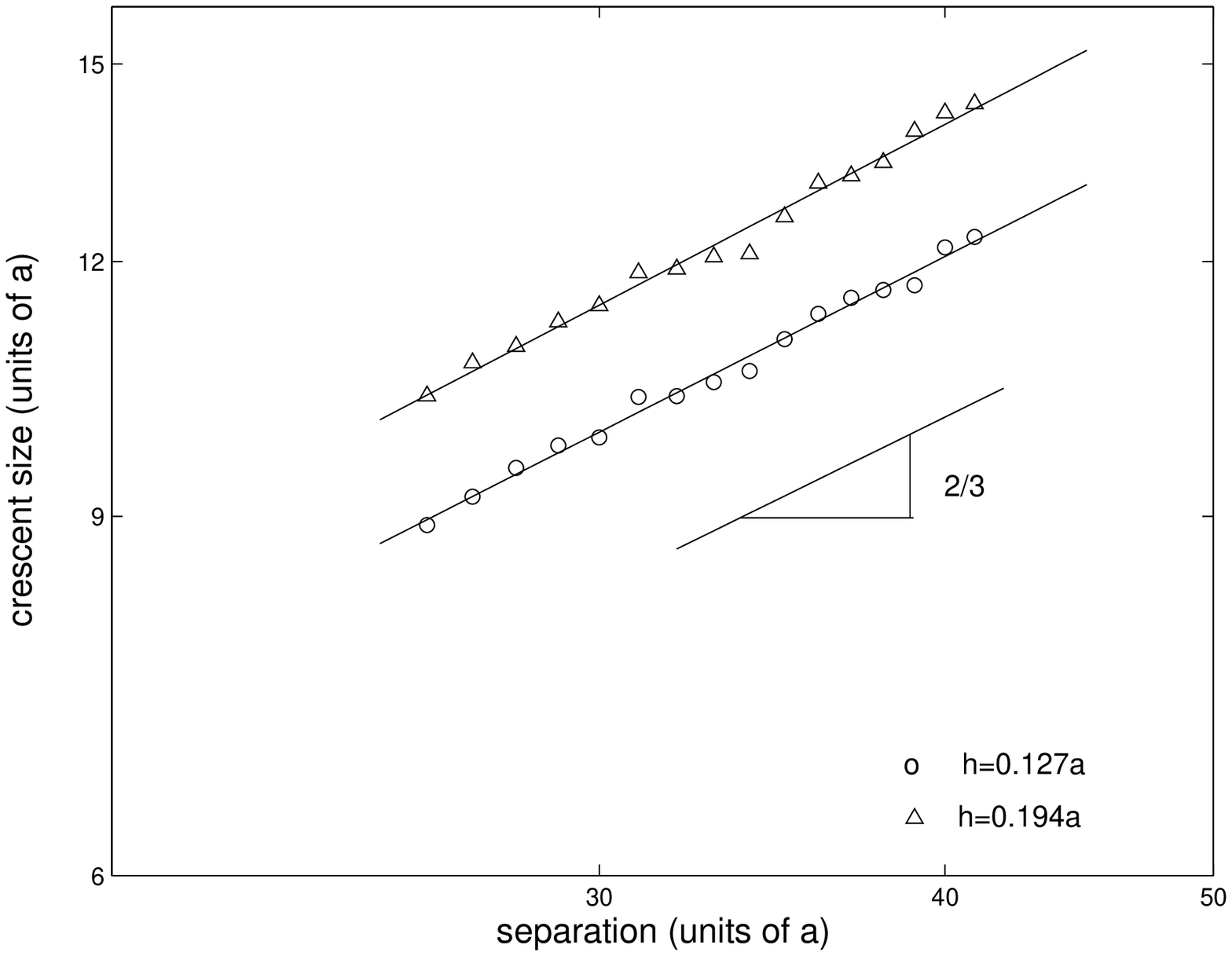}
\caption{Plots of crescent size versus separation for two different thicknesses. Crescent size is measured from the same simulated shape as in Fig.~\ref{width_sep}. The fitted lines have slope 0.689 and 0.709.}
\label{crescent_sep}
\end{center}
\end{figure}

Fig.~\ref{crescent_sep} shows the plot of crescent size versus separation for the same simulated shapes. Again, we observe a scaling relationship between crescent size and separation, and it is different from width scaling. The fitted values of slope are 0.689 and 0.709, close to 2/3. 

\begin{figure}[!htb]
\begin{center}
\includegraphics[width=0.8\textwidth]{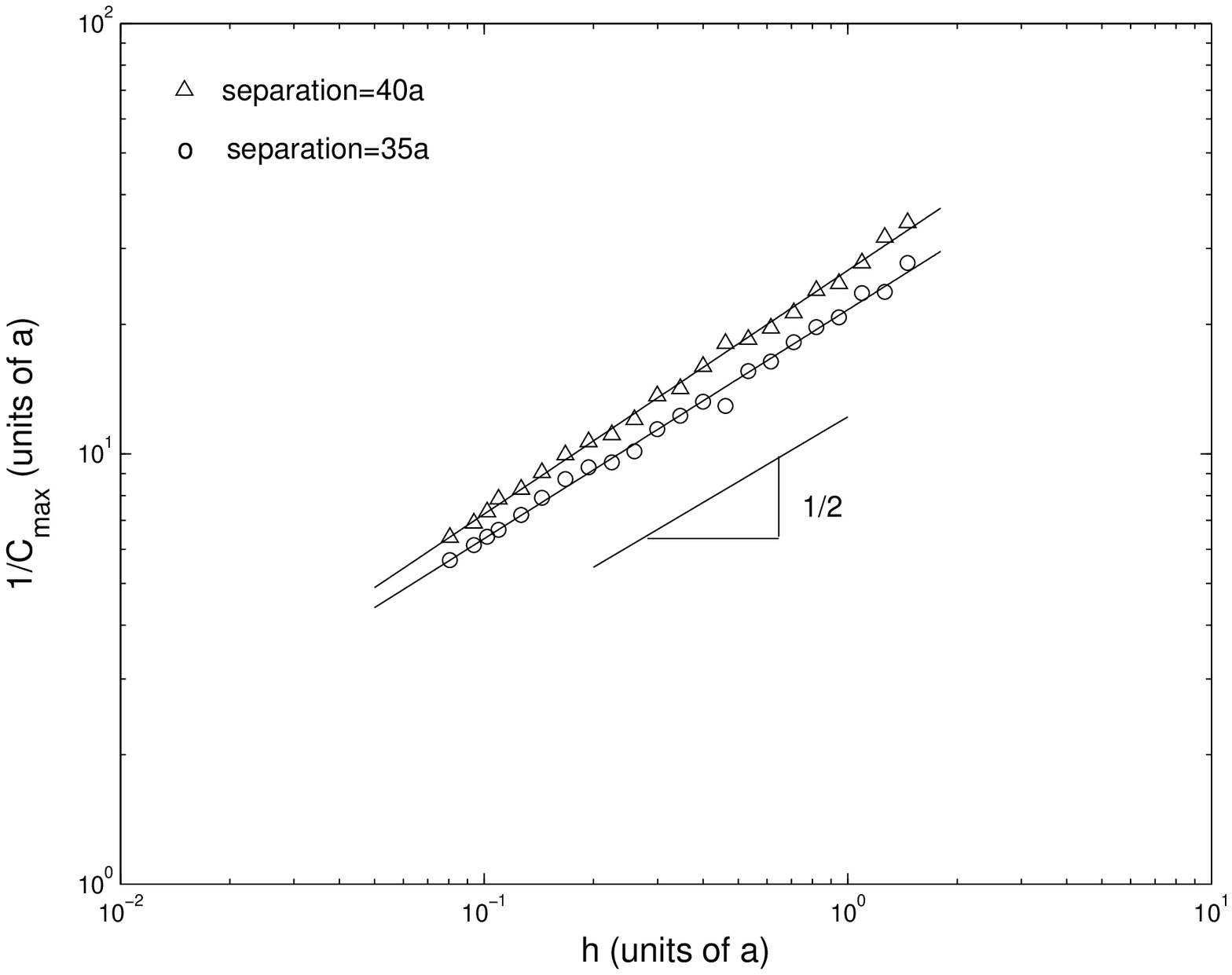}
\caption{Plots of width versus thickness with separation fixed at $35a$ and $40a$. The fitted lines have slope of 0.532 and 0.567, for separation of $35a$ and $40a$, respectively.}
\label{width_thickness_check}
\end{center}
\end{figure}

Having obtained the scaling law on separation, we want to check our thickness scaling of width for other values of separation. In IV.B, we obtain the thickness scaling when separation is fixed at $30a$. We now go back and go through the same procedure of changing the thickness but with separation fixed at other values. Fig.~\ref{width_thickness_check} displays the plots of width versus thickness with separation fixed at $35a$ and $40a$. The values of the fitted slope are 0.532 and 0.567, both close to 1/2. This is consistent with our results obtained earlier.  

The scaling observations of crescent width remind us of the properties of a ring ridge, formed by pushing against a convex shell until a dent is formed \cite{pogorelov} \cite{witten1}. If the shell is thin and elastic, the boundary of the ring ridge becomes arbitrarily sharply curved compared to the radius of the dent, and the ring ridge has both stretching and bending energies, like a straight ridge. More quantitatively, in the limit of zero thickness, the ring ridge becomes a sharp circular crease of radius $R_s$. For physical sheet with finite thickness $h$, the ring ridge posesses a nonzero radius of curvature $w$, and the circular ridge line must expand by an amount proportional to $w$, thus creating a strain of the order $w/R_s$. The stretching energy in the ridge region is of order $S \approx (\kappa/h^2)(w/R_s)^2 w R_s \approx \kappa w^3 /(h^2 R_s)$. The bending energy is of order $B \approx \kappa (1/w)^2 w R_s  \approx \kappa R_s/w$. The optimal $w$ is the one that minimizes the total energy: $w \sim (hR_s)^{1/2}$. Although our observation shows that crescent width scales with thickness $h$ to the $1/2$ power, seemingly similar to the scaling property of a ring ridge, it is fundamentally different from it because the crease radius $R_s$ in our system changes with thickness $h$. More specifically, as $R_s$ is similar to the crescent size, which changes as $h^{1/3}R^{2/3}$ according to our observation, the ring ridge scaling suggests $w \sim h^{1/2}(h^{1/3}R^{2/3})^{1/2} \sim h^{2/3}R^{1/3}$. This is different what we observed for width scaling.

\section{Discussion and Conclusion}
In this paper, we have explored numerically the properties of an isolated core/crescent region, especially its scaling properties. The crescent region is formed by applying six forces to the sheet in a controlled way, such that it is similar to the crescent region in a $d$ cone. It has two characteristic length scales: one is the width of the crescent region, and the other is its radius of curvature. Through our numerical investigation, we found that the width of the crescent region scales with thickness to the 1/2 power and with separation to the 1/2 power. This scaling property is different from a straight ridge, which scales with thickness to the $1/3$ power, and is also different from a ring ridge, which scales with thickness to the $2/3$ power for our system, as shown in Section IV.C. The scaling exponent lies between those of a straight ridge and a ring ridge. Although this structure is energetically more expensive than straight ridge, we do not see any evidence that it tends to break up into straight ridges. This fact indicates that the energy gain from the formation of straight ridges must be less than the energy cost of eliminating the crescent-like region. We note that this $h^{1/2}$ scaling of the crescent width contrasts with the behavior seen in a related crescent experiment by Das \textit{et al} \cite{maha-cond}. 

On the other hand, the radius of curvature of the crescent is revealed to scale with thickness to the 1/3 power and with separation to the 2/3 power. This observation is consistent with what we observed for the crescent in a $d$ cone. The confining ring has the same position as the separation between the potentials in the scaling law. This seems reasonable; they both represent the length scale at which we confine the sheet to produce the desired crescent.  

It is worth noting that deformation angle $\beta$ in our structure is directly related to the deformation $\epsilon$ in a $d$ cone. Using the shape of $d$ cone calculated when assuming there is no stretching energy \cite{cerda-prl}, we found that $\beta$ and $\epsilon$ are related by $\beta = \pi - 4.82 \epsilon$. A typical value of $\beta=2.7$ rad in our study corresponds to $\epsilon= 0.09$ of $d$ cone, which is a typical value in our previous study of $d$ cone \cite{last}. More specifically, if we write $R_c \approx G (hS^2)^{1/3}$ for our isolated crescent region, where $R_c$ is the crescent size and $S$ is the separation, then it is found that $G = 2.88 \pm 0.31$ for the typical value of $\beta=2.7$ rad. In comparison, the value of $A$ for a $d$ cone cresent region is $G = 1.54 \pm 0.12$ for the typical value of $\epsilon=0.09$ \cite{last}. We note that the isolated crescent region size is almost twice of a $d$ cone crescent size, while the sheet in both cases has same side length and experiences similar deformations. This is consistent with what we expect: we wanted to isolate the crescent region so that it is possible to study it in greater details. In addition, Cerda \textit{et al} \cite{cerda-nature} propose that crescent size of $d$ cone has a scaling relationship with $\epsilon$. However, in our simulation, we do not observe any meaningful scaling relationship between crescent size and $\pi-\beta$. Neither for width and $\pi-\beta$. The reason may be attributed to the fact that we do not have a large range of $\beta$. 

Our investigation of this isolated crescent region shows that it exhibits two different features: one as a crescent in $d$ cone; the other as a structure with scaling property between a ring ridge and a straight ridge. Our work helps to confirm the puzzling scaling law observed earlier, and reveal some interesting properties about the crescent region that would not be possible to obtain for a $d$ cone. In the light of this study, the puzzle we proposed in Section I is not yet completely clarified. It is our hope that the knowledge gained in this study can be helpful in the clarification of the puzzle. In future work, variable lattice spacing technique could be used to improve the numerical accuracy in the regions where curvature is large. Other variants of the crescent region could be constructed and studied.

\begin{acknowledgments}
The author sincerely thanks Prof. Tom Witten for innumerable advice and inspiring discussions. This paper is part of author's doctoral research done under his supervision. This work was supported in part by the National Science Foundation's MRSEC Program under Grant Number DMR-0213745.
\end{acknowledgments}


\begin{thebibliography}{99}

\bibitem{nelson1} D.R. Nelson, L. Peliti, J. Phys. (Paris) {\bf 48}, 1085 (1987). 

\bibitem{alex3} A.E. Lobkovsky, S. Gentes, H. Li, D. Morse, T.A. Witten, Science \textbf{270}, 1482 (1995).

\bibitem{kramer1} E.M. Kramer, T.A. Witten, Phys. Rev. Lett. {\bf 78}, 1303 (1997).

\bibitem{sackmann} E. Sackmann, P. Eggl, C.Fahn, H. Bader, H. Ringsdorf, M. Schollmeier, Ber. Bunsenges. Phys. Chem. {\bf 89}, 1198 (1985).

\bibitem{spector} M.S. Spector, E. Naranjo, S. Chiruvolu, J.A. Zasadzinski, Phys. Rev. Lett. {\bf 73}, 2867 (1994).

\bibitem{chianelli} R.R. Chianelli, E.B. Prestrige, T.A. Pecoraro, J.P. DeNeufville, Science {\bf 203}, 1105 (1979).

\bibitem{bull} A.J. Bull, Geol. Mag. {\bf 69}, 73 (1932).

\bibitem{stut} P. Stutenkemper, R. Brasche, in \textit{Occupant Protection in Frontal Impacts}, Proceedings of the Seventh International Technical Conference on Experimental Safety Vehicles, Paris, 1979 (U.S. Department of Transportation, Wahsington, D.C., 1980).

\bibitem{newton} R.E. Newton, in \textit{Shock and Vibration Handbook}, edited by C.M. Harris and C.E. Crede (McGraw-Hill, New York, 1988).

\bibitem{cerda-nature} E. Cerda, S. Chaieb, F. Melo, L. Mahadevan, Nature {\bf 401}, 46 (1999).

\bibitem{witten} T.A. Witten, H. Li, Europhys. Lett. {\bf 23}, 51 (1993).

\bibitem{alex1} A.E. Lobkovsky, Phys. Rev. E \textbf{53}, 3750 (1996).

\bibitem{alex2} A.E. Lobkovsky, T.A. Witten, Phys. Rev. E \textbf{55}, 1577 (1997).

\bibitem{brian} B.A. DiDonna, Phys. Rev. E \textbf{66}, 016601 (2002).

\bibitem{brian2} B.A. DiDonna, T.A. Witten, Phys. Rev. Lett. {\bf 87}, 206105 (2001).

\bibitem{ben-amar} M. Ben Amar, Y. Pomeau, P. Roy. Soc. Lond. A {\bf 453}, 729 (1997).

\bibitem{chaieb1} S. Chaieb, F. Melo, J.-C. Geminard, Phys. Rev. Lett. \textbf{80}, 2354 (1998).

\bibitem{cerda-prl} E. Cerda, L. Mahadevan, Phys. Rev. Lett. {\bf 80}, 2358 (1998).

\bibitem{kramer} E.M. Kramer, J. Math. Phys. {\bf 38}, 830 (1997).

\bibitem{brian1} B.A. DiDonna, T.A. Witten, S.C. Venkataramani, E.M. Kramer, Phys. Rev. E {\bf 65}, 016603 (2001).

\bibitem{venka} S.C. Venkataramani, T.A. Witten, E.M. Kramer, R.P. Geroch, J. Math. Phys. {\bf 41}, 5107 (2000).

\bibitem{boud} A. Boudaoud, P. Patricio, Y. Couder, M. Ben Amar, Nature \textbf{407}, 718 (2000).

\bibitem{chaieb2} S. Chaieb, F. Melo, Phys. Rev. E \textbf{60}, 6091 (1999).

\bibitem{boud1} T. Mora, A. Boudaoud, Europhys. Lett. \textbf{59}, 41 (2002).

\bibitem{last} T. Liang, T.A. Witten, Phys. Rev. E {\bf 71}, 016612 (2005).

\bibitem{last1} T. Liang, T.A. Witten, Phys. Rev. E {\bf 73}, 046604 (2006).

\bibitem{maha-new} E. Cerda, L. Mahadevan, Proc. R. Soc. A, {\bf 461}, 671 (2005).

\bibitem{ajay} A. Gopinathan, T.A. Witten, S.C. Venkataramani, Phys. Rev. E {\bf 65}, 036613 (2002). 

\bibitem{ortiz} M. Ortiz, G. Gioia, J. Mech. Phys. Solids {\bf 42}, 531 (1994). 

\bibitem{audoly} B. Audoly, Phys. Rev. Lett. {\bf 83}, 4124 (1999).

\bibitem{keer} L.M. Keer, M.A.G. Silva, J. Appl. Mech., 1121 (1972).

\bibitem{boud2} A. Boudaoud, S. Chaieb, Phys. Rev. E {\bf 64}, 050601(R) (2001).

\bibitem{bha} K. Bhattacharya, R.D. James, J. Mech. Phys. Solids {\bf 47}, 531 (1999).

\bibitem{cuccia} L.A. Cuccia, R.B. Lennox, Am. Chem. Soc. {\bf 208}, 50 (1994). 

\bibitem{nelson} H.S. Seung, D.R. Nelson, Phys. Rev. A \textbf{38}, 1005 (1988).

\bibitem{mansfield} E.H. Mansfield, \textit{The Bending and Stretching of Plates} (Pergamon, New York, 1964).

\bibitem{landau} L.D. Landau, E.M. Lifshitz, \textit{Theory of Elasticity} (Pergamon Press, New York, 1986).

\bibitem{recipe} W.H. Press, S.A. Teukolsky, W.T. Vetterling, B.P. Flannery, \textit{Numerical Recipe in C} (Cambridge University Press, Cambridge, 1996).

\bibitem{struik} D.J. Struik, \textit{Lectures on Classical Differential Geometry} (Addison-Wesley Publishing Company, Massachusetts, 1961).

\bibitem{milman} R.S. Milman, G.D. Parker, \textit{Elements of Differential Geometry} (Prentice Hall, Englewood Cliffs, NJ, 1977).

\bibitem{witten1} T.A. Witten, Rev. Mod. Phys. {\bf 79}, 643 (2007).

\bibitem{maha-cond} M. Das, A. Vaziri, A. Kudrolli, L. Mahadevan, Phys. Rev. Lett. {\bf 98}, 014301 (2007).

\bibitem{pogorelov} A.V. Pogorelov, 1960, as summarized in Ref.[37] Section 15.

\end{thebibliography}
\end{document}